\documentclass{article} 
\usepackage[final]{colm2026_conference}

\usepackage{microtype}
\usepackage{hyperref}
\usepackage{url}
\usepackage{booktabs}

\usepackage{natbib}

\usepackage{lineno}

\definecolor{darkblue}{rgb}{0, 0, 0.5}
\hypersetup{colorlinks=true, citecolor=darkblue, linkcolor=darkblue, urlcolor=darkblue}

\usepackage[utf8]{inputenc}
\usepackage[T1]{fontenc}
\usepackage[most]{tcolorbox}
\definecolor{customblue}{HTML}{243885}

\newcommand{\myparatight}[1]{\smallskip\noindent{\bf {#1}:}~}

\usepackage{pifont}

\usepackage{amsthm}
\usepackage{amsmath}
 
\usepackage{amssymb}
\usepackage{float}
\usepackage{graphics}
\usepackage{graphicx}

\usepackage{algorithm}
\usepackage{algpseudocode}

\usepackage{bm}
\usepackage{color}
\usepackage{multirow}
\usepackage{makecell}
\usepackage{subfig}
\usepackage{soul}

\usepackage{bigstrut,multirow,rotating}
\usepackage{tabto}

\usepackage{booktabs}
\usepackage{enumitem}
\usepackage{bbm}
\usepackage{dsfont}

\usepackage{diagbox}
\usepackage{cancel}
\usepackage{tabularx}
\usepackage{threeparttable}

\usepackage{balance}

\hypersetup{
	colorlinks=true,
	linkcolor=blue,
	filecolor=magenta,      
	urlcolor=cyan,
	pdftitle={Overleaf Example},
	pdfpagemode=FullScreen,
}

\usepackage{color, colortbl}
\definecolor{greyL}{RGB}{230,248,255}

\allowdisplaybreaks

\definecolor{PineGreen}{RGB}{0,139,114}
\definecolor{BrickRed}{RGB}{140,55,62}

\usepackage{xspace}

\newcommand{\alg}{{\textsf{AgentLocate}}\xspace}

\newtheorem{definition}{Definition}

\definecolor{ffe8e7L}{RGB}{255,255,248}

\algnewcommand\algorithmicforparallel{\textbf{for}}
\algnewcommand\algorithmicdoinparallel{\textbf{do in parallel}}
\algdef{S}[FOR]{ForParallel}[1]{%
  \algorithmicforparallel\ #1\ \algorithmicdoinparallel}

\title{Who Broke the System? Failure Localization in LLM-Based Multi-Agent Systems}

\author{Yufei Xia$^1$, Anjun Gao$^1$, Yueyang Quan$^2$, Zhuqing Liu$^2$, Minghong Fang$^1$\\
$^1$University of Louisville, $^2$University of North Texas \\
}

\begin{document}
\ifcolmsubmission
\linenumbers
\fi

\maketitle

\begin{abstract}

Large language model (LLM) based multi-agent systems enable complex problem solving through coordinated reasoning and action, but their distributed structure also introduces new challenges in diagnosing system-level failures. When an execution fails, identifying which agent is responsible and at what point the trajectory first becomes irreversibly misdirected is difficult due to long-horizon interactions and tightly coupled agent behaviors. In this paper, we study the problem of failure localization in LLM-based multi-agent systems and present \alg, a framework that attributes failures to both a specific agent and the earliest decisive step. \alg combines an LLM-based judging mechanism with multi-perspective verification by independent evaluators, whose assessments are aggregated using a confidence-aware strategy. The resulting feedback is further used to adapt the judge through lightweight fine-tuning, improving attribution quality. We evaluate \alg on two complementary benchmarks covering diverse tasks, agent configurations, and trajectory lengths. Experimental results show that \alg consistently outperforms existing failure localization methods in identifying both responsible agents and failure steps, while remaining efficient in terms of token usage and running time.

\end{abstract}


\section{Introduction} 
\label{sec:intro}

Large language model (LLM) based multi-agent systems~\citep{li2024survey,hong2024metagpt,li2023camel,wu2024autogen,han2024llm,talebirad2023multi} have recently gained prominence as an effective framework for tackling problems that exceed the capabilities of a single language-model-driven agent. 
By distributing responsibilities across specialized components, these systems enable richer forms of reasoning, more flexible planning, and coordinated workflows that span diverse domains. They have been applied to areas such as software engineering~\citep{qian2024chatdev,tufano2024autodev}, scientific discovery~\citep{boiko2023emergent,ferraro2025generative}, information gathering~\citep{shen2023hugginggpt,sun2025multi}, and complex web-based tasks~\citep{zhou2023webarena,mialon2023gaia}, where collaborative agent behaviors can yield stronger performance than isolated models. This growing adoption highlights the potential of multi-agent architectures as a foundation for building increasingly capable AI systems.

Despite these advantages, the increasing adoption of LLM-based multi-agent systems has also revealed a growing set of reliability concerns. The very features that make multi-agent architectures powerful, such as distributed decision making, role specialization, and tool-mediated interactions, also create new pathways for failures to arise~\citep{zhang2025agent,cemri2025multi,kong2025aegis}. Unlike single-agent pipelines, where errors typically stem from a localized misprediction, multi-agent systems introduce interdependent behaviors in which a subtle mistake by one component can ripple across subsequent steps, distort shared state, or derail collaborative planning. 
These intertwined execution paths make it difficult to determine not only \emph{when} a failure occurs but also \emph{which} agent or interaction first pushed the system off course. As multi-agent workflows become more intricate and are deployed in increasingly realistic environments, understanding and diagnosing these failure dynamics becomes essential for ensuring dependable operation.

Automatic failure localization in LLM-based multi-agent systems has recently attracted growing attention, with several methods~\citep{zhang2025agentracer,zhang2025agent,banerjee2025did,kong2025aegis} proposed to attribute failures to a responsible agent and a decisive step. However, our empirical results suggest that existing approaches remain only partially effective, largely because multi-agent failures emerge from long-horizon, intertwined execution paths where causal responsibility is diffuse and difficult to isolate. For example, counterfactual replay methods like AgenTracer~\citep{zhang2025agentracer} are closer to causal testing in principle, yet their conclusions can be unstable in multi-agent settings because altering one step often changes subsequent prompts, tool outcomes, and coordination patterns, making the failure-inducing action hard to pin down consistently across variants. In contrast, template- or taxonomy-driven methods like AEGIS~\citep{kong2025aegis} infer agent responsibility by matching behaviors to predefined error patterns, but this reliance on curated categories limits coverage for emergent reasoning and coordination breakdowns that do not align cleanly with pre-specified error patterns. 
Another straightforward solution is to apply existing poisoning forensics approaches~\citep{zhang2025taught,zhang2025traceback} to failure localization in multi-agent systems. 
However, as demonstrated in our later experiments, these poisoning forensics approaches are not effective in localizing failures in multi-agent systems because they are tailored to attacks that often leave detectable traces even when engineered to remain stealthy, whereas failures in multi-agent systems typically emerge from incremental missteps and coordination drift, leaving no clear traceable origin.

\noindent
\textbf{Our contributions: }%
In this paper, we aim to bridge this gap by enabling reliable failure localization in LLM-based multi-agent systems. We propose \alg, a practical framework that pinpoints both the responsible agent and the step at which the trajectory first becomes decisively misdirected.
The core motivation is that a single attribution pass is often brittle for long, tool-mediated trajectories, so localization should be treated as a process that can be challenged, verified, and improved rather than a one-shot decision. Concretely, \alg begins by having an LLM Judge produce an explicit hypothesis about the failure location, and it supports both an all-at-once inspection of the full trajectory and a step-by-step protocol that examines the growing prefix to surface the earliest decisive point as soon as sufficient evidence appears.  This design makes the method compatible with real multi-agent logs, while keeping the overall pipeline simple enough to run repeatedly during debugging.

\alg then turns this initial hypothesis into a more reliable localization through a Judge-Evaluator refinement stage. After the Judge proposes a candidate agent-step pair, multiple independent Evaluators re-analyze the same trajectory conditioned on that hypothesis, each producing its own predicted location along with a rationale and a self-reported confidence.  \alg aggregates these assessments via a confidence-weighted voting scheme, yielding a verified estimate that is less sensitive to the quirks of any single model run or prompt.  Importantly, the Evaluators are prompted with diverse styles (e.g., base, concise, and evidence-focused) to encourage complementary reasoning behaviors during verification. Finally, \alg converts the Judge prediction, evaluator feedback, and the aggregated decision into fine-tuning instances and adapts the Judge via parameter-efficient fine-tuning, so that the Judge incorporates the evaluators’ critiques as supervision, improving localization fidelity.

We summarize our main contributions in this paper as follows:

\begin{list}{\labelitemi}{\leftmargin=1em \itemindent=0em \itemsep=0.3em \parsep=0em \topsep=0em \partopsep=0em}

\item 
We introduce \alg, a practical framework for failure localization in LLM-based multi-agent systems that identifies both the responsible agent and the earliest decisive step leading to system-level failure.

\item 
We design a Judge-Evaluator refinement mechanism that verifies and aggregates independent localization hypotheses using confidence-aware voting, and leverages evaluator feedback to improve the Judge via parameter-efficient fine-tuning.

\item 
We perform extensive evaluations on two representative benchmarks, Who\&When and Aegis-Bench, showing that \alg consistently outperforms existing failure localization and poisoning forensics approaches across different models, datasets, and evaluation settings, while remaining efficient in token usage and running time.

\end{list}	


\section{Preliminaries and related work} \label{sec:related}

\subsection{LLM-based multi-agent systems: a primer} 
\label{sec:related_Primer}

We consider an LLM-based multi-agent system $\mathcal{M}$ whose agents jointly respond to a user query $q$.
Let $\mathcal{N} = \{1,\dots,N\}$ be the index set of agents. Following~\citet{zhang2025agent,cemri2025multi}, we adopt a turn-based protocol, i.e., the interaction unfolds in discrete decision steps, and at each step only a single agent is allowed to act. 
The multi-agent system $\mathcal{M}$ is formalized as $\mathcal{M} = \langle \mathcal{N}, \mathcal{S}, \{\mathcal{A}_i\}_{i \in \mathcal{N}}, \mathcal{F}, \rho \rangle$,
where $\mathcal{S}$ is the collection of admissible states, $\mathcal{A}_i$ is the action space of agent $i$, $\mathcal{F}$ is the environment transition function, and $\rho$ is a scheduler that specifies which agent acts at each decision step. 
Formally, we take
$
\rho: \{0,1,2,\dots\} \to \mathcal{N},
$
so that $\rho(t)$ denotes the index of the agent that is allowed to act at time $t$. If the active agent at time $t$ is $\rho(t)$ and it takes action $a_t \in \mathcal{A}_{\rho(t)}$ in state $s_t$, then the next state is drawn from $\mathcal{F}(\cdot \mid s_t, a_t, \rho(t))$.

Each agent $i \in \mathcal{N}$ follows a policy $\pi_i$ that maps the observed state, the user query, and a portion of the past interaction to an action. Let $\mathcal{H}_t \subseteq \{a_0,\dots,a_{t-1}\}$ denote the action history visible at step $t$, where $a_0$ is the initial action; then the action taken at time $t$ by the active agent $\rho(t)$ can be represented as
$
a_t = \pi_{\rho(t)}(s_t, \mathcal{H}_t, q).
$
We formalize the overall trajectory produced by the multi-agent system $\mathcal{M}$ in responding to the user query $q$ as $\tau = (s_0, a_0, s_1, a_1, \dots, s_T)$,
where $s_0$ is the initial state, and $T$ denotes the step at which execution terminates, either because a predefined stopping condition is met or because a terminal state is encountered.

\subsection{Related work} 
\label{sec:related_work}

\myparatight{LLM-based multi-agent systems}%
LLM-based multi-agent systems~\citep{li2024survey,hong2024metagpt,li2023camel,wu2024autogen,han2024llm,talebirad2023multi} have emerged as a powerful approach for solving complex tasks through coordinated reasoning and action among multiple agents~\citep{wu2024autogen,li2023camel,qian2024chatdev}. Prior work studies a range of coordination mechanisms, including role specialization~\citep{talebirad2023multi}, structured communication~\citep{li2023camel,qian2024chatdev}, deliberative interactions~\citep{du2023improving}, and adaptive agent topologies~\citep{zhuge2024gptswarm,hu2024automated}, as well as the integration of external tools for web interaction~\citep{zhou2023webarena}, code execution~\citep{wang2024executable}, and file manipulation~\citep{chen2024sheetagent}. Existing systems span different levels of automation, from manually designed frameworks~\citep{wu2024autogen} to partially~\citep{chen2025optimizing,yue2025masrouter,yuksekgonul2024textgrad,zhuge2024gptswarm} and fully automated agentic pipelines~\citep{zhang2025evoflow,nie2025weak,gao2025flowreasoner}. Despite steady progress, system-level failures~\citep{zhu2025llm,zhang2025agent,kong2025aegis} can still occur due to miscoordination or compounding inaccuracies across agents, motivating the need for methods that improve reliability and analysis in diverse multi-agent settings.

\myparatight{Failure localization for LLM-based multi-agent systems}%
Failure localization for LLM-based multi-agent systems has gained attention as these systems grow more complex, introducing system-level failures through multi-agent coordination, tool use, and structured interactions. 
Recent studies~\citep{zhang2025agentracer,zhang2025agent,banerjee2025did,kong2025aegis} frame automated failure localization as identifying the responsible agent and the earliest decisive step in an execution trajectory, a task that remains challenging even for strong reasoning models. 
Representative baselines include AgenTracer~\citep{zhang2025agentracer}, which uses counterfactual trajectory replay to localize responsibility, and AEGIS~\citep{kong2025aegis}, which relies on curated error patterns for agent-level localization. Despite their progress, existing methods still face limitations in localization accuracy and efficiency, motivating further research on precise failure attribution for complex multi-agent systems.

Note that poisoning forensics~\citep{shan2022poison,cheng2023beagle,hammoudeh2022identifying,jia2024tracing,rose2024utrace,zhang2025taught,zhang2025traceback,jia2025promptlocate,perry2019no,gao2026patching} addresses a distinct problem, focusing on detecting adversarial manipulations such as corrupted training data or maliciously injected retrieval content that induce incorrect model behavior. In contrast, failure localization aims to identify execution-time failures that arise naturally during system operation. A detailed comparison is provided in Appendix~\ref{sec:appendix_poison}.

\section{The \alg algorithm} \label{sec:defense}

Let $t^{\star}(\tau)$ denote the \emph{decisive failure step} (the earliest step at which a single corrected action would reverse the failure outcome) and let $i^{\star}(\tau)$ denote the \emph{failure-responsible agent} scheduled at that step. The full definition of these quantities is provided in Appendix~\ref{sec:appendix_problem}. We now formally define the multi-agent systems failure localization problem below.
\begin{definition}[Multi-agent systems failure localization problem]
    \label{definition1}
Given a query $q$ and an observed failed trajectory $\tau$ that does not successfully fulfill $q$, the goal of multi-agent systems failure localization is to infer the pair $(i^{\star}(\tau),\, t^{\star}(\tau))$, which identifies the agent and the earliest step whose misaction is decisive for the system-level failure.
\end{definition}
To address this problem, we propose \alg, a framework that performs reliable failure localization through a three-stage iterative cycle of hypothesis generation, multi-perspective verification, and adaptive refinement.
To infer $(i^{\star}, t^{\star})$ reliably,  \alg treats localization as a verifiable and improvable process rather than a one-shot decision. An LLM-based Judge first hypothesizes both the responsible agent and the earliest decisive step that causes system-level failure. A set of independent Evaluators with diverse reasoning styles then re-examine the same trajectory, each producing an alternative prediction with a rationale and confidence score. \alg consolidates these via confidence-weighted voting and feeds the resulting assessments back to refine the Judge through parameter-efficient fine-tuning. Through this Judge-Evaluator refinement cycle, \alg achieves accurate failure localization across diverse multi-agent settings.

\subsection{Decisive failure hypothesis generation}

As noted in prior work~\citep{zhang2025agent,kong2025aegis,banerjee2025did}, attributing failures in multi-agent systems requires identifying subtle causal dependencies between actions of heterogeneous agents and the global outcome. The annotation burden for humans is substantial, and agreement among annotators is often low. The LLM-as-a-judge paradigm~\citep{zheng2023judging,kocmi2023large,chiang2023can,li2025generation,son2024llm}, developed in recent evaluation research, demonstrates that LLMs can provide structured assessments of complex behaviors by synthesizing long-range dependencies and applying flexible reasoning patterns.

In this work, we operationalize failure localization with an LLM-based Judge that reasons over the logged trajectory and the user query to infer the responsible agent and the earliest decisive step. Let $p$ denote a prompt specifying a reasoning protocol for interpreting a trajectory $\tau$ under a decisive-error criterion. Concretely, we seek a timestep $\hat{t}$ such that replacing the action at $\hat{t}$ with an ideal action would reverse the failure outcome, i.e., the corrected trajectory $\mathfrak{R}(\tau,\hat{t})$ succeeds on query $q$.
Because $\mathfrak{R}$ involves an idealized counterfactual intervention, the Judge does not execute $\mathfrak{R}$ directly. Instead, it approximates this criterion by reasoning over the logged trajectory and the query, and returns the step and corresponding acting agent that it believes best satisfies the counterfactual reversal property. 
With $\hat{i}=\rho(\hat{t})$ as in Eq.~(\ref{idea_agent}) (in Appendix~\ref{sec:appendix_problem}), the hypothesis generation procedure is:
\begin{align}
\label{hypothesis_gen}
	(\hat{i},\hat{t}) = \text{Judge}(p,\tau, q).
\end{align}

Since trajectories may span many steps, the reasoning protocol can expose the sequence to the Judge in different ways.
In the all-at-once setting~\citep{zhang2025agent}, the Judge is given the full trajectory $(s_{0},a_{0},s_{1},a_{1},\ldots,s_{T})$ together with the query and must directly identify both the agent responsible for the failure and the timestep at which the decisive error occurred.
A complementary option is the step-by-step setting~\citep{zhang2025agent}, where the Judge observes only the growing prefix $(s_{0},a_{0},\ldots,s_{t},a_{t})$ at each timestep $t$.
After each partial reveal, the Judge evaluates whether the current action already satisfies the counterfactual criterion above. If so, it outputs $(\rho(t),t)$; if not, it continues to the next step. This incremental process enables the Judge to identify the earliest decisive event as soon as sufficient evidence appears in the trajectory.
The prompts $p$ used for all-at-once and step-by-step are provided in Appendix~\ref{Prompts_step_by_step}.
This module produces a single hypothesis $(\hat{i},\hat{t})$ based solely on the Judge's internal reasoning. This dependence motivates an external mechanism that can independently assess, critique, and, when necessary, correct the hypothesis, which leads to the second component of our method.

\subsection{Evaluator verification}
\label{Verification_model}

The decisive failure hypothesis generation module provides a tentative hypothesis $(\hat{i},\hat{t})$ for the decisive failure location based on the Judge's internal reasoning. The goal of the evaluator verification module is to scrutinize this hypothesis using multiple independent views of the same trajectory, and to transform the single Judge output into a robust, aggregated label that can subsequently guide adaptive improvement.

To this end, \alg instantiates a set of Evaluators indexed by $h\in\mathcal{H}$. Each Evaluator receives $(\tau,q)$ together with the Judge's hypothesis $(\hat{i},\hat{t})$. 
Evaluator $h\in\mathcal{H}$ is conditioned on its own prompt and reasoning protocol, such as base, concise, or evidence-focused styles. The prompts for these three protocols are shown in Appendix~\ref{Prompts style for Evaluator}. The evaluator then re-analyzes the failure and proposes the agent-step pair that it believes best satisfies the decisive-error criterion, without being forced to agree with the Judge.
Formally:
\begin{align}
\label{Evaluator_each}
	(i_h,t_h,r_h,c_h) = \text{Evaluator}_h(e_h,\tau,q,\hat{i},\hat{t}),
\end{align}
where $(i_h,t_h)$ is the decisive failure location predicted by Evaluator $h$, $r_h$ is a natural-language justification derived from its reasoning process, $c_h \in [0,1]$ reflects its self-reported confidence in this assessment,
and $e_h$ is the prompt used by Evaluator $h$, such as the Base, Concise, or Evidence prompts defined above.
Passing $(\hat{i},\hat{t})$ as an input allows Evaluators to explicitly comment on or depart from the Judge's proposal, yielding explanations that directly highlight agreement or disagreement with the initial inference.

To consolidate the $H$ independent judgments, \alg defines a confidence-weighted support function over candidate agent-step locations:
\begin{align}
	A(i,t) = \sum_{h \in \mathcal{H}} c_h \mathbf{1}\{(i_h,t_h) = (i,t)\}.
\end{align}
The verification module then selects the agent-step pair with the highest aggregate support as its final estimate:
\begin{align}
\label{Evaluator_agg}
	(i_{\text{eval}}, t_{\text{eval}}) = \arg\max_{(i,t)} A(i,t).
\end{align}

This stage therefore produces two complementary outputs: (1) an evaluator-aggregated decisive failure location $(i_{\text{eval}}, t_{\text{eval}})$, and (2) the set of rationales and confidence scores $\{(r_h,c_h)\}_{h \in \mathcal{H}}$ that explain and qualify the aggregate decision. The adaptive Judge improvement module uses both the comparison between $(\hat{i},\hat{t})$ and $(i_{\text{eval}}, t_{\text{eval}})$ and the accompanying evaluator feedback to construct training signals that refine the Judge's future inferences.
Our experiments show that treating the Judge’s hypothesis $(\hat{i}, \hat{t})$ or the Evaluators’ aggregated output $(i_{\text{eval}}, t_{\text{eval}})$  as the final output leads to poor localization accuracy.

\subsection{Adaptive Judge improvement}

Because the decisive failure hypothesis generation module depends entirely on the Judge's internal reasoning process, systematic deviations between the Judge's predictions and the causal definition of decisive failure would persist unless the Judge is updated accordingly. The verification module in Section~\ref{Verification_model} exposes such deviations and provides explanations indicating where the Judge's reasoning diverges from the decisive-error criterion.
If $(\hat{i},\hat{t}) = (i_{\mathrm{eval}}, t_{\mathrm{eval}})$, then the Evaluators' analyses support the Judge's hypothesis, and their rationales can still reveal missing evidence, unclear causal links, or overly late attribution that can be corrected in future reasoning. Conversely, if $(\hat{i},\hat{t}) \neq (i_{\mathrm{eval}}, t_{\mathrm{eval}})$, the discrepancy highlights a misalignment between the Judge's assessment and the trajectory's causal structure. In both cases, the adaptive improvement mechanism uses the evaluator feedback to refine the Judge and reduce future reasoning errors.

To translate these signals into systematic model improvement, we construct explicit training instances that encode the Judge's prediction, the Evaluators' assessments, and the aggregated decisive failure location. Specifically, for query $q$, \alg constructs an instance:
\begin{align}
	z_q = \{\tau, q, \hat{i},\hat{t}, \{i_h,t_h,r_h,c_h\}_{h \in \mathcal{H}}, i_{\mathrm{eval}}, t_{\mathrm{eval}} \}.
\end{align}

The fine-tuning dataset is then defined as
$
\mathcal{D} = \{z_q\}_{q \in \mathcal{Q}},
$
which collects instances from a designated refinement set $\mathcal{Q}$ of failed trajectories that is kept disjoint from the trajectories used for evaluation. After constructing $\mathcal{D}$, we fine-tune the Judge on these instances with LoRA~\citep{hu2022lora} to better align its reasoning with the causal definition of decisive failure. Let $\text{Judge}_{\text{ft}}$ denote the resulting fine-tuned Judge. For any query $q$ with trajectory $\tau$, including previously unseen queries at evaluation time, the final decisive failure location is obtained by:
\begin{align}
\label{final_res}
(i_{\text{final}}, t_{\text{final}}) = \text{Judge}_{\text{ft}}(p,\tau,q).
\end{align}

This adaptive improvement module establishes a feedback-driven refinement cycle: Decisive Failure Hypothesis Generation produces a hypothesis, Evaluator Verification critiques and aggregates external assessments, and Adaptive Judge Improvement updates the Judge to promote increasingly accurate and causally aligned failure attribution over time. The cycle may be repeated on successive batches of failures, improving the Judge as more verified instances are accumulated.

Algorithm~\ref{alg:main} in the Appendix shows the pseudocode of our proposed \alg algorithm. For each failed query $q \in \mathcal{Q}$ with trajectory $\tau_q$, the algorithm first invokes the Judge to generate an initial failure hypothesis $(\hat{i}, \hat{t})$. Conditioned on this hypothesis, each Evaluator $h \in \mathcal{H}$ independently produces a candidate failure location $(i_h, t_h)$, together with a rationale $r_h$ and a confidence score $c_h$. These evaluator outputs are combined using a confidence-weighted aggregation mechanism to obtain a verified failure location $(i_{\text{eval}}, t_{\text{eval}})$. The Judge hypothesis, evaluator feedback, and aggregated result are then packaged into a refinement instance $z_q$, and the collection of all such instances forms the fine-tuning dataset $\mathcal{D}$. 
Fine-tuning the Judge on $\mathcal{D}$ yields $\mathrm{Judge}_{\mathrm{ft}}$, which is then used to localize failures for any query at inference time via Eq.~(\ref{final_res}).


\section{Experiments} \label{sec:exp}

\subsection{Experimental setup}

\myparatight{Datasets and comparison methods}%
We evaluate on two benchmark datasets: \textbf{Who\&When}~\citep{zhang2025agent}, comprising Algorithm-Generated and Hand-Crafted subsets, and \textbf{Aegis-Bench}~\citep{kong2025aegis}. For Who\&When, we use 50\% for training, 10\% for validation, and 40\% for testing; for Aegis-Bench, we follow the original setting~\citep{kong2025aegis} with 600 samples for testing and the remaining 8,933 for training. In our method, refinement uses failed training trajectories, and evaluation uses the test split. Full dataset details are provided in Appendix~\ref{sec:appendix_datasets}. We compare \alg against four failure-localization baselines: WhichAgent~\citep{zhang2025agent}, AgenTracer~\citep{zhang2025agentracer}, ECHO~\citep{banerjee2025did}, and AEGIS~\citep{kong2025aegis}. We additionally compare against two poisoning forensics methods, RAGOrigin~\citep{zhang2025taught} and RAGForensics~\citep{zhang2025traceback}, in Section~\ref{sec:discussion_limitation}. Full descriptions of all baselines are provided in Appendix~\ref{sec:appendix_baselines}.

\myparatight{Evaluation metrics and parameter settings}%
For Who\&When, we report \textbf{agent-level accuracy} (proportion of failures with the responsible agent correctly identified) and \textbf{step-level accuracy} (proportion with the decisive error step correctly localized). For Aegis-Bench, which lacks step-level annotations, we report \textbf{agent-level accuracy} and \textbf{pair-level accuracy} (both the faulty agent and its predefined error mode correctly identified). The Judge is instantiated as one of Qwen2.5-7B-Instruct~\citep{qwen7b2025} (Qwen-7B), Llama-3.1-8B-Instruct~\citep{grattafiori2024llama} (Llama-8B), Mistral-7B-Instruct-v0.3~\citep{mistral7b2023} (Mistral-7B), or GPT-4o~\citep{achiam2023gpt}, and evaluated under both an all-at-once mode and a step-by-step mode (prompts in Appendix~\ref{Prompts_all_once}); all experiments are conducted without access to ground-truth answers. Three Evaluators with distinct prompt styles (base, concise, evidence; see Appendix~\ref{Prompts style for Evaluator}) review each Judge output, using the same underlying LLM as the Judge. The Judge is fine-tuned via LoRA~\citep{hu2022lora} with rank 64, scaling factor 128, dropout 0.05, 3 epochs, effective batch size 16. GPT-4o is fine-tuned via the OpenAI platform with equivalent settings. Unless otherwise specified, we use Qwen-7B in the all-at-once mode, and code will be released upon acceptance.

\subsection{Experimental results}

\begin{table*}[t]
\centering
\small
\begin{tabular}{l lcccc}
\toprule
& & \multicolumn{2}{c}{All-at-once} & \multicolumn{2}{c}{Step-by-step} \\
\cmidrule(lr){3-4} \cmidrule(lr){5-6}
Model & Method & Agent-level & Step-level & Agent-level & Step-level \\
\midrule
\multirow{5}{*}{Qwen-7B}
& WhichAgent  & 30.95 & 19.05 & 7.14  & 2.38  \\
& AgenTracer  & 45.24 & 11.90 & 38.10 & 7.14  \\
& ECHO        & 30.95 & 14.29 & 9.52  & 2.38  \\
& AEGIS       & 33.33 & --    & 23.81 & --    \\
\rowcolor{greyL}
\cellcolor{white} & \alg & 69.05 & 38.10 & 69.05 & 33.33 \\
\midrule
\multirow{5}{*}{Llama-8B}
& WhichAgent  & 45.24 & 23.81 & 28.57 & 9.52  \\
& AgenTracer  & 40.48 & 2.38  & 45.24 & 7.14  \\
& ECHO        & 35.71 & 11.90 & 23.81 & 14.29 \\
& AEGIS       & 38.10 & --    & 50.00 & --    \\
\rowcolor{greyL}
\cellcolor{white} & \alg & 54.76 & 23.81 & 54.76 & 16.67 \\
\midrule
\multirow{5}{*}{Mistral-7B}
& WhichAgent  & 47.62 & 16.67 & 7.14  & 4.76  \\
& AgenTracer  & 45.24 & 11.90 & 35.71 & 9.52  \\
& ECHO        & 40.48 & 11.90 & 30.95 & 9.52  \\
& AEGIS       & 21.43 & --    & 26.19 & --    \\
\rowcolor{greyL}
\cellcolor{white} & \alg & 50.00 & 26.19 & 38.10 & 16.67 \\
\midrule
\multirow{5}{*}{GPT-4o}
& WhichAgent  & 52.38 & 11.90 & 28.57 & 19.05 \\
& AgenTracer  & 45.24 & 11.90 & 61.90 & 9.52  \\
& ECHO        & 40.48 & 14.29 & 19.05 & 9.52  \\
& AEGIS       & 23.81 & --    & 21.43 & --    \\
\rowcolor{greyL}
\cellcolor{white} & \alg & 61.90 & 30.95 & 64.29 & 33.33 \\
\bottomrule
\end{tabular}
\caption{Localization performance on the Who\&When \textit{Algorithm-Generated} subset. Higher agent-level and step-level accuracy (\%) indicate better localization performance.}
\label{result_alg}
\vspace{-.10in}
\end{table*}

\myparatight{Our proposed \alg is effective}%
Table~\ref{result_alg} reports the localization performance of \alg on the Algorithm-Generated subset under all-at-once and step-by-step protocols, compared against four baselines across four LLMs. \alg achieves the strongest overall performance compared with all baselines at both agent-level and step-level accuracy across all model and evaluation configurations, reaching average agent-level accuracy above 50\%. Notably, while WhichAgent and AgenTracer show competitive agent-level performance in certain settings, their step-level accuracy degrades substantially in the step-by-step protocol. AEGIS, limited to agent-level attribution, exhibits instability across models. \alg maintains stable and generalizable localization quality across all settings. Results on the more complex Hand-Crafted subset and on Aegis-Bench, where \alg similarly achieves the strongest agent-level performance, are provided in Table~\ref{result_hand} and Table~\ref{aegis_result} (Appendix) with detailed discussion in Appendix~\ref{sec:appendix_handcrafted}.

\myparatight{How does the failure propagate through the multi-agent system?}%
Fig.~\ref{fig:Failure propagation} (Appendix) presents an illustrative case of how failure emerges and propagates in a multi-agent system, drawn from the Algorithm-Generated subset of the Who\&When dataset. A user requests an estimate of bottle-collecting earnings using Wikipedia. Three domain specialists (an algorithm design expert, a recycling expert, and a validation expert) collaborate with a computer terminal to handle the query. The algorithm design expert first outlines the solution steps. From Step~\ding{183} to Step~\ding{186}, the recycling expert fails to retrieve bottle pricing data from Wikipedia. The earliest decisive failure occurs at Step~\ding{187}, where the validation expert mistakenly uses an assumed price rather than the extracted value. Although the validation expert later flags the need to recheck Wikipedia at Step~\ding{190}, the terminal concludes the interaction based on the earlier incorrect assumption, yielding a wrong final output. This example illustrates how a single erroneous decision at Step~\ding{187} redirects the entire trajectory toward failure, underscoring the importance of pinpointing the earliest decisive error and the agent responsible for it.

\myparatight{Granular analysis of step-level localization}%
Fig.~\ref{fig:step_difference} (Appendix) measures the deviation between predicted and true failure steps under Qwen-7B (all-at-once). \alg yields the most predictions within small deviation ranges on both subsets (most of the cases within 0 to 2 steps on Algorithm-Generated) and the fewest in large deviation ranges on the more challenging Hand-Crafted subset. We discuss in detail in Appendix~\ref{sec:appendix_step_deviation}.

\myparatight{Step-level accuracy under different tolerances}%
Fig.~\ref{fig:step_tolerance} (Appendix) evaluates tolerance-based step-level accuracy under Qwen-7B (all-at-once), counting a prediction as correct if it falls within a given step tolerance window of the true failure step. \alg achieves the highest accuracy across all tolerance values on both subsets, with a particularly clear advantage on the Hand-Crafted subset as the window widens. Details are in Appendix~\ref{sec:appendix_step_tolerance}.

\myparatight{Impact of trajectory length on localization accuracy}%
We analyze localization accuracy across four token-length categories under Qwen-7B (all-at-once). As shown in Tables~\ref{dataset length level-agent} and~\ref{dataset length level-step} (Appendix), \alg remains the strongest overall method across length ranges on both subsets, showing no clear monotonic degradation on Algorithm-Generated and retaining the strongest relative performance on Hand-Crafted trajectories, where other methods deteriorate. Detailed results and discussion are in Appendix~\ref{sec:appendix_token_length}.

\myparatight{\alg is computationally efficient and cost-effective}%
 Tables~\ref{tab:token_cost_tokens}, \ref{tab:token_cost_money} and \ref{tab:token_cost_time} (Appendix) compare token usage, monetary cost and running time under Qwen-7B, with training overhead included for both \alg and AgenTracer. \alg achieves efficiency comparable to training-free baselines (e.g., 451K vs.\ 249K tokens for WhichAgent on Algorithm-Generated, all-at-once) while running over $20\times$ faster than AgenTracer (315 vs.\ 6,752 seconds on Hand-Crafted, all-at-once) at a fraction of the cost (\$0.148 vs.\ \$3.204), demonstrating a clearly superior efficiency-performance trade-off.

\myparatight{Impact of different LLMs used as Evaluators}%
Table~\ref{results with different evaluator model}  (Appendix) presents the performance of \alg when the Judge is fixed to Qwen-7B and the Evaluators vary among Qwen-7B, Llama-8B, Mistral-7B, and GPT-4o. The differences across evaluator models are modest. For instance, in the Algorithm-Generated and all-at-once mode, agent-level accuracy remains within a narrow range of 52\% to 69\%, and step-level accuracy falls between 23\% and 38\%. A similar degree of stability is observed on the Hand-Crafted dataset.

\myparatight{Impact of the number of Evaluators}%
Table~\ref{results with different evaluator number} (Appendix) reports agent-level and step-level accuracy under Qwen-7B (Algorithm-Generated) as the number of Evaluators varies from 1 to 7. Increasing from 1 to 3 Evaluators yields a notable accuracy gain (e.g., 35.71\%$\to$69.05\% for agent-level accuracy under all-at-once mode), while further increasing to 5 or 7 brings no consistent improvement. Three Evaluators thus provide the best trade-off between performance and efficiency.

\myparatight{Impact of refinement rounds}%
Tables~\ref{results_loop_algorithm} and~\ref{results_loop_handcrafted} (Appendix) report localization accuracy under 1 to 3 Judge-Evaluator refinement rounds on both subsets. Core performance is already established after round 1 (e.g., Qwen-7B achieves 69.05\% agent-level on Algorithm-Generated, all-at-once), with additional rounds yielding only marginal and inconsistent gains. A single refinement round is thus sufficient in practice.


\section{Discussion} 
\label{sec:discussion_limitation}

\myparatight{Different variants of \alg}%
Table~\ref{results with variants} (Appendix) reports an ablation analysis of three variants of our proposed \alg. Variant I directly adopts the output of the Decisive Failure Hypothesis Generation stage as the final prediction, using $(\hat{i}, \hat{t})$ without any further verification or refinement. Variant II relies solely on the Evaluator Verification stage and outputs the aggregated evaluator decision $(i_{\text{eval}}, t_{\text{eval}})$ as the final result. Variant III removes the iterative feedback mechanism and instead fine-tunes the Judge directly on the original benchmark data, such as the Who\&When dataset.
As shown in Table~\ref{results with variants} (Appendix), the complete \alg consistently outperforms all three variants, indicating that each component of the framework contributes to strong failure localization performance. In particular, the relatively strong performance of Variant II indicates not only the effectiveness of multi-Evaluator diversity aggregation itself, but also the high quality of the Evaluator-generated supervision later used for Judge adaptation.
{Although Variant III is trained directly on the original benchmark trajectories with ground-truth labels $(i^{\star}(\tau), t^{\star}(\tau))$, its performance still remains below that of the full \alg. A likely reason is that one-shot fine-tuning on final labels provides only limited supervision, especially for long trajectories, since it offers little learning signal beyond the final answer. 
In contrast, \alg adopts a verify-and-refine framework that converts failed trajectories into Evaluator-enriched training instances and adapts the Judge accordingly, supplying richer corrective signals that support reliable localization.

\myparatight{Comparison between \alg and poisoning forensics approaches}%
Table~\ref{results with RAG} (Appendix) compares \alg against RAGOrigin~\citep{zhang2025taught} and RAGForensics~\citep{zhang2025traceback} on the Algorithm-Generated subset. Both forensics methods perform substantially worse, with agent-level accuracy below 34\% under both evaluation modes, compared to \alg's 69.05\%. This gap stems from a fundamental mismatch: poisoning forensics methods are designed to detect adversarial perturbations that leave identifiable traces, whereas failures in multi-agent systems arise from benign yet complex interactions, subtle reasoning errors or coordination breakdowns that produce no such detectable signatures. These results confirm that failure localization requires dedicated methods beyond repurposed forensics approaches. Full method descriptions and analysis are in Appendix~\ref{sec:appendix_forensics}.

\myparatight{Analysis of frequently mislocalized agents}%
Table~\ref{Top 3 error agent} (Appendix) reports the agents most frequently mislocalized by different methods under the Qwen-7B model in the all-at-once mode. Across methods and datasets, several agents repeatedly appear among the most frequently mislocalized ones, such as \texttt{Verification\_Expert} in the Algorithm-Generated dataset and \texttt{WebSurfer} in the Hand-Crafted dataset. Fig.~\ref{fig:error agent freq} (Appendix) shows that these agents are also among the most common true failure sources.
This overlap suggests a close connection between high-frequency failure agents and frequent mis-attribution targets. These agents typically occupy late-stage verification or retrieval roles, where upstream mistakes first become visible, causing the Judge to misattribute downstream symptoms as root causes. This visibility bias is amplified in longer trajectories. A natural future direction for \alg is to model error propagation more explicitly, separating the true upstream source from the point where the failure surfaces.


\section{Conclusion} 
\label{sec:conclusion}

We study failure localization in LLM-based multi-agent systems, where system-level errors arise from complex, long-horizon interactions among multiple agents. We present \alg, a Judge-Evaluator framework that localizes failures by identifying the responsible agent and the earliest decisive step in a failed execution. By treating localization as a verifiable process rather than a one-shot decision, \alg combines hypothesis generation, multi-perspective verification, confidence-aware aggregation, and evaluator-guided Judge refinement. Experiments on two complementary benchmarks show that \alg achieves stronger localization accuracy than existing attribution and forensics-based methods, while remaining efficient in token usage, cost, and running time.

\section*{Acknowledgments}

We thank the reviewers for their constructive comments.

\section*{Ethics Statement}

This work investigates failure localization in LLM-based multi-agent systems to improve the reliability, transparency, and debuggability of multi-agent workflows by identifying the responsible agent and the earliest decisive step associated with a system-level failure. Our framework follows the LLM-as-a-judge paradigm: an LLM-based Judge analyzes a failed trajectory and the user query to infer the failure location, while independent Evaluators re-examine the same trajectory for verification and refinement.
We envision this approach being used in research and development settings involving real multi-agent logs, long-horizon interactions, and debugging workflows, particularly as such systems are applied to software engineering, scientific discovery, and web-based tasks.

\bibliographystyle{colm2026_conference}
\bibliography{refs}


\newpage
\appendix

\begin{algorithm}[t]
	\caption{\alg.}
	\label{alg:main}
	\renewcommand{\algorithmicrequire} {\textbf{Input:}}
	\renewcommand{\algorithmicensure}{\textbf{Output:}} 
	\begin{algorithmic}[1]
		\Require Failed refinement queries $\mathcal{Q}$ with trajectory $\tau_q$ for each $q \in \mathcal{Q}$; prompt $p$; the set of Evaluators $\mathcal{H}$; prompt $e_h$ for each Evaluator $h \in \mathcal{H}$.
		\Ensure Fine-tuned Judge $\text{Judge}_{\text{ft}}$
		\State Initialize fine-tuning dataset $\mathcal{D} \leftarrow \emptyset$
		\For{each failed query $q \in \mathcal{Q}$}
         \State Obtain Judge hypothesis $(\hat{i}, \hat{t}) \leftarrow \text{Judge}(p, \tau_q, q)$
		\For{each Evaluator $h$ in $\mathcal{H}$}
				\State Obtain $(i_h, t_h, r_h, c_h) \leftarrow \text{Evaluator}_h(e_h, \tau_q, q, \hat{i}, \hat{t})$
		\EndFor
		\State Compute $(i_{\text{eval}}, t_{\text{eval}})$ via confidence-weighted voting (Eq.~(\ref{Evaluator_agg}))
		\State Construct the fine-tuning instance
		$z_q = \{\tau_q, q, \hat{i}, \hat{t}, \{i_h, t_h, r_h, c_h\}_{h \in \mathcal{H}}, i_{\text{eval}}, t_{\text{eval}}\}$
		\State $\mathcal{D} \gets \mathcal{D} \cup \{z_q\}$
		\EndFor
		\State Fine-tune Judge on $\mathcal{D}$ to obtain $\text{Judge}_{\text{ft}}$
	    \State \Return $\text{Judge}_{\text{ft}}$
	\end{algorithmic}
\end{algorithm}

\section{Comparison between poisoning forensics and failure localization}
\label{sec:appendix_poison}

Poisoning forensics research is mainly concerned with uncovering adversarial manipulations that deliberately induce incorrect model behavior, such as poisoned training examples or maliciously injected retrieval documents. In these scenarios, the observed errors are the direct result of an attacker’s intervention, and the forensic objective is to attribute the undesired behavior to a small set of harmful sources. A substantial body of work has explored this problem~\citep{shan2022poison,cheng2023beagle,hammoudeh2022identifying,jia2024tracing,rose2024utrace,zhang2025taught,zhang2025traceback,jia2025promptlocate,perry2019no,gao2026patching}, typically under the assumption that adversarial actions leave behind identifiable statistical anomalies, causal footprints, or behavioral signatures that remain observable after deployment.
Early studies focus on traditional machine learning settings. For instance, prior work~\citep{shan2022poison} introduces an iterative clustering-and-pruning strategy that progressively removes benign training samples until the remaining subset isolates the poisoned data responsible for the attack. Other efforts extend poisoning forensics to distributed environments; for example, work such as~\citet{jia2024tracing} addresses forensic analysis in federated learning systems, where malicious clients may inject poisoned updates during collaborative training. 
More recently, poisoning forensics has been adapted to LLM pipelines, particularly retrieval-augmented generation. 
Methods proposed in~\citet{zhang2025taught,zhang2025traceback} demonstrate that it is possible to efficiently identify poisoned documents that poison retrieval results and ultimately lead to adversarial outputs.

Failure localization, by contrast, addresses a fundamentally different class of errors. Rather than adversarial manipulation, its goal is to diagnose execution-time breakdowns that arise naturally during system operation. In complex LLM-based systems, errors may accumulate through sequential decision making, imperfect reasoning, or interactions among system components, and the final incorrect outcome often results from compounding effects rather than a single corrupted input. As a result, responsibility cannot be determined by isolating an anomalous data source or artifact alone. Consequently, methods designed for poisoning attribution do not perform well when applied to failure localization, as they are not suited to identifying the critical decision point or component where the system first deviates from correct behavior. Our experimental results further confirm this limitation, motivating the need for dedicated techniques specifically designed for failure localization.

\section{Problem statement}
\label{sec:appendix_problem}

From Section~\ref{sec:related_Primer}, let $\tau$ denote the complete trajectory generated by $\mathcal{M}$ in response to query $q$. In practice, trajectories can fail due to incorrect intermediate reasoning, miscoordination among agents, or unstable action selection; we say $\tau$ \emph{succeeds} if $\mathcal{M}$ fulfills $q$, and \emph{fails} otherwise.

To investigate why a trajectory fails, we seek to locate the earliest time step whose incorrect decision initiates the failure, assuming access to an oracle that provides the ideal action at each step. For each time index $t \in \{0,\dots,T-1\}$, let $a_t^{\star}$ denote the ideal action that the active agent $\rho(t)$ should have taken at time $t$. Given a trajectory $\tau=(s_0,a_0,\dots,s_T)$ and any index $t$, define $\mathfrak{R}(\tau,t)$ as the counterfactual trajectory constructed as follows: the prefix $(s_0,a_0,\dots,s_t)$ is kept fixed except that $a_t$ is replaced by $a_t^{\star}$; from state $s_t$ onward, the environment and agents are re-simulated under the same policies and scheduler, but now beginning from this corrected action.
For a failed trajectory, we define the set of time indices whose single-step correction changes the outcome from failure to success as:
\begin{align}
    \mathcal{K}(\tau)= \{ t \in \{0,\dots,T-1\} \mid \tau \text{ fails and } \mathfrak{R}(\tau,t) \text{ succeeds} \}.
\end{align}
Whenever $\mathcal{K}(\tau)$ is nonempty, the earliest such step serves as the trajectory's \emph{decisive failure step}:
\begin{align}
    t^{\star}(\tau)= \min \mathcal{K}(\tau),
\end{align}
and the \emph{failure-responsible agent} is the agent scheduled at that earliest step:
\begin{align}
    \label{idea_agent}
    i^{\star}(\tau)= \rho(t^{\star}(\tau)).
\end{align}
These quantities provide a principled notion of which agent and which specific decision, in the earliest-step sense, are causally responsible for the observed failure.

\section{Prompts for all-at-once and step-by-step modes}
\label{Prompts_all_once}
\begin{tcolorbox}[colback=gray!10,
                  colframe=customblue!80,
                  width=\linewidth,
                  arc=1mm, auto outer arc,
                  boxrule=1pt,
                  left=1mm, right=1mm, top=0.1mm, bottom=0.1mm,
                  fontupper=\footnotesize,
                  fonttitle=\footnotesize,
                  title = Prompt for all-at-once
                 ]
You are an AI assistant tasked with analyzing a multi-agent conversation history when solving a real world problem.\\
The problem is: \textcolor{red}{\{problem\}}.\\[0.5em]
Identify which agent made an error, at which step, and explain the reason for the error.\\
Here’s the conversation: \textcolor{red}{\{failure log\}}\\[0.75em]
Based on this conversation, please predict the following:\\
1.\ The name of the agent who made a mistake that should be directly responsible for the wrong solution to the real world problem. If there are no agents that make obvious mistakes, decide one single agent in your mind. Directly output the name of the Expert.\\
2.\ At which step the mistaken agent first made a mistake. For example, in a conversation structured as follows:\\[0.25em]
\{\texttt{"agent a": "xx",}\\
\texttt{"agent b": "xxxx",}\\
\texttt{"agent c": "xxxxx",}\\
\texttt{"agent a": "xxxxxxx"}\}\\[0.25em]
each entry represents a `step' where an agent provides input. The `x' symbolizes the speech of each agent. If the mistake is in agent c's speech, the step number is 2. If the second speech by `agent a' contains the mistake, the step number is 3, and so on. Please determine the step number where the first mistake occurred.\\
3.\ The reason for your prediction.\\[0.5em]
Please answer in the format:\\
Agent Name: (Your prediction)\\
Step Number: (Your prediction)\\
Reason for Mistake: (Your reason)
\end{tcolorbox}

\begin{tcolorbox}[colback=gray!10,
                  colframe=customblue!80,
                  width=\linewidth,
                  arc=1mm, auto outer arc,
                  boxrule=1pt,
                  left=1mm, right=1mm, top=0.1mm, bottom=0.1mm,
                  fontupper=\footnotesize,
                  fonttitle=\footnotesize,
                  title = Prompt for step-by-step
                 ]
You are an AI assistant tasked with evaluating the correctness of each step in an ongoing multi-agent conversation aimed at solving a real-world problem.\\
The problem being addressed is: \textcolor{red}{\{problem\}}.\\[0.5em]
Here is the conversation history up to the current step:\\
\textcolor{red}{\{conversation history\}}\\[0.5em]
The most recent step (\textcolor{red}{\{step index\}}) was by \textcolor{red}{\{agent name\}}.\\[0.5em]
Your task is to determine whether this most recent agent's action (Step \textcolor{red}{\{step index\}}) contains an error that could hinder the problem-solving process or lead to an incorrect solution. Please respond with ``Yes'' or ``No'' and provide a clear explanation for your judgment.\\[0.5em]
Note: Please avoid being overly critical in your evaluation. Focus on errors that clearly derail the process.\\[0.5em]
Attention: Respond ONLY in the format:\\
1.\ Yes/No.\\
2.\ Reason: [Your explanation here]
\end{tcolorbox}
\label{Prompts_step_by_step}

\section{Prompts for Evaluator}

\label{Prompts style for Evaluator}
\begin{tcolorbox}[colback=gray!10,
                  colframe=customblue!80,
                  width=\linewidth,
                  arc=1mm, auto outer arc,
                  boxrule=1pt,
                  left=1mm, right=1mm, top=0.1mm, bottom=0.1mm,
                  fontupper=\footnotesize,
                  fonttitle=\footnotesize,
                  title = Prompt for base
                 ]
\textbf{System:} You are a neutral failure-attribution judge for a multi-agent LLM system.\\
\textbf{Role:} Given a predicted responsible agent, step and a numbered failure log, decide whether the prediction is correct and cite supporting or contradicting evidence by line numbers.\\
\textbf{Format:} Respond ONLY with a single JSON object that strictly follows the required schema; do not include any additional prose or explanations.
\end{tcolorbox}

\begin{tcolorbox}[colback=gray!10,
                  colframe=customblue!80,
                  width=\linewidth,
                  arc=1mm, auto outer arc,
                  boxrule=1pt,
                  left=1mm, right=1mm, top=0.1mm, bottom=0.1mm,
                  fontupper=\footnotesize,
                  fonttitle=\footnotesize,
                  title =  Prompt for concise
                 ]
\textbf{System:} You are a terse, accuracy-focused JSON evaluator.\\
\textbf{Role:} Check whether the predicted responsible agent and step is justified by the numbered dialogue and encode your verdict and brief rationale in JSON fields.\\
\textbf{Format:} Output exactly one valid JSON object and nothing else, no preface, no code fences, no extra text.
\end{tcolorbox}

\begin{tcolorbox}[colback=gray!10,
                  colframe=customblue!80,
                  width=\linewidth,
                  arc=1mm, auto outer arc,
                  boxrule=1pt,
                  left=1mm, right=1mm, top=0.1mm, bottom=0.1mm,
                  fontupper=\footnotesize,
                  fonttitle=\footnotesize,
                  title =  Prompt for evidence
                 ]
\textbf{System:} You are a rigorous adjudicator who prioritizes evidence before making any judgment.\\
\textbf{Role:} First inspect the numbered failure log to collect key supporting and opposing lines, then determine whether the predicted responsible agent and step is correct based on that evidence.\\
\textbf{Format:} Return a single JSON object containing both the evidence references and the final verdict; the reply must be valid JSON only.
\end{tcolorbox}

\section{Details of datasets}
\label{sec:appendix_datasets}
In our experiments, we evaluate our approach alongside various baseline methods on two benchmark datasets.
It is worth noting that failure localization in multi-agent systems is an emerging research topic, and the availability of benchmark datasets remains limited.

\myparatight{Who\&When~\citep{zhang2025agent}}%
The Who\&When dataset is a dedicated benchmark for attributing failures within multi-agent systems. 
For every failure case, the dataset records the posed task, the user query, the entire dialogue and interaction history among agents, the correct answer to the task, and a detailed annotation specifying which agent caused the failure and at which step the error originated.
The dataset is organized into two separate subsets described below.

\begin{list}{\labelitemi}{\leftmargin=1.5em \itemindent=-0.3em \itemsep=.2em  \topsep=1pt}
\item Algorithm-Generated: This subset includes 191 distinct agents distributed across 126 failure cases, with trajectories averaging 8.6 decision steps.

\item Hand-Crafted: This subset comprises 5 distinct agents over 58 failure cases, with an average trajectory length of 50 steps.
\end{list}

\myparatight{Aegis-Bench~\citep{kong2025aegis}}%
The Aegis-Bench dataset contains 9,533 failure trajectories across six task domains, namely MATH, SciBench, GSM8K, HumanEval, MMLU, and GAIA, and six multi-agent system frameworks, including LLM Debate, MacNet, AgentVerse, Dylan, SmolAgents, and Magentic-One. Overall, it includes 24,843 injected error instances, capturing broad diversity in tasks, agent structures, and failure characteristics.

For the Who\&When datasets, we apply 60\% of the data for Judge model development, including 50\% for training and 10\% for validation, and reserve the remaining 40\% for testing. To ensure a balanced split, the training and test sets are matched in terms of average trajectory length and complexity. In the Algorithm-Generated subsets, both splits contain an average of 8.7 decision steps. In the Hand-Crafted subsets, the training and test sets contain an average of 49.8 and 54.2 decision steps, respectively. For Aegis-Bench, we follow the original setting in~\cite{kong2025aegis}, using 600 samples from six task domains for testing and the remaining 8,933 samples for training.

\section{Comparison methods}
\label{sec:appendix_baselines}

\myparatight{WhichAgent~\citep{zhang2025agent}}%
It assigns responsibility by locating the earliest point at which a multi-agent trajectory departs from the correct course of action. A judge model inspects the agents’ decisions, either globally or at individual steps, and determines which agent and which step introduced the first meaningful deviation from the intended solution path, thus pinpointing the source of the failure.

\myparatight{AgenTracer~\citep{zhang2025agentracer}}%
It determines which agents cause a failure by replaying trajectories with systematic modifications and observing how these alterations impact the final outcome. By assembling a tailored dataset and training a lightweight tracing model that incorporates counterfactual signals across multiple levels of granularity, the approach yields accurate identification of the agent at fault and the exact step where the error originates.

\myparatight{ECHO ~\citep{banerjee2025did}}%
ECHO method employs a hierarchical context framework that incrementally condenses multi-agent trajectories, enabling the model to represent fine-grained decision mistakes as well as broader coordination patterns. This structured encoding is paired with a set of heterogeneous evaluators whose outputs are integrated through a confidence-weighted consensus procedure.

\myparatight{AEGIS~\citep{kong2025aegis}}%
It performs attribution using a curated collection of trajectories in which particular error categories and responsible agents are artificially introduced and meticulously annotated. At test time, the model contrasts the observed agent actions with these predefined error templates, allowing it to infer the error type and identify the agent whose behavior most closely matches the injected failure pattern.

\myparatight{RAGOrigin~\citep{zhang2025taught}}%
It uncovers harmful data by examining how each retrieved document influences the model’s produced answer. By selectively modifying or suppressing individual passages and observing the resulting changes in the LLM’s output, the method pinpoints the documents whose information most directly leads to an incorrect or malicious response.

\myparatight{RAGForensics~\citep{zhang2025traceback}}%
It carries out traceback by repeatedly retrieving a set of candidate documents and using an LLM classifier to determine whether each one contributes to the incorrect answer. Documents identified as poisoning sources are removed, and the procedure continues until only benign materials remain, thereby isolating the specific entries responsible for the erroneous response.

\section{Extended results on other datasets}
\label{sec:appendix_handcrafted}
\myparatight{Performance on Hand-Crafted dataset}%
Table~\ref{result_hand} reports localization performance on the Hand-Crafted subset, where trajectories average 50 steps and can exceed 100, posing substantially greater difficulty than the Algorithm-Generated subset. \alg continues to outperform all baselines at the agent level across all four LLMs and both evaluation modes, demonstrating robustness under long-horizon settings. For GPT-4o, adaptation is limited to OpenAI's built-in supervised fine-tuning framework rather than LoRA, yet \alg still achieves competitive agent-level accuracy, indicating that the multi-Evaluator-constructed dataset transfers effectively even under constrained adaptation conditions.

For step-level accuracy, all methods struggle more on this subset due to the increased trajectory length, which dilutes the failure signal across lengthy interaction histories. Nevertheless, \alg produces the largest proportion of predictions within small deviation ranges and the fewest within large deviation ranges among all compared methods, confirming its stronger step-localization capability in long-horizon and intricate failure scenarios. WhichAgent and AgenTracer exhibit marked performance drops in the step-by-step setting, while ECHO remains consistently low across both modes. These results collectively indicate that \alg delivers the most reliable failure attribution on complex multi-agent trajectories.

\myparatight{Performance on Aegis-Bench dataset}%
Table~\ref{aegis_result} provides an additional evaluation of failure localization on the Aegis-Bench~\citep{kong2025aegis} dataset, using two evaluation modes across three models: Qwen-7B, Llama-8B, and Mistral-7B. Note that in the Aegis-Bench, the objective is to identify the agent responsible for the failure and then determine the specific error mode the agent made, based on a predefined taxonomy containing 14 distinct error modes. All baselines are applied to the Aegis-Bench dataset except for ECHO, which relies on step-based evaluation and cannot be configured appropriately for this benchmark. The operational pipeline of \alg remains unchanged in this experiment. Because the evaluation must both identify the correct agent and assign the correct error mode, this dual requirement increases task difficulty and leads to generally lower pair-level accuracy.
In terms of overall performance, \alg continues to outperform the competing methods at the agent-identification level, achieving average agent-level accuracy values above 50\% when paired with Qwen-7B and Llama-8B. However, the gains at the pair-level  accuracy are more modest, with smaller margins relative to some baselines. These findings suggest that \alg preserves its effectiveness in related but more challenging error attribution settings, showing resilience in identifying failure sources even when finer-grained error mode categorization is required. Notably, \alg is primarily designed to localize failures in terms of Who and When for exact agent and step, rather than to distinguish fine-grained error modes, which partly explains its limited gains on pair-level accuracy.

\section{Granular analysis of step-level localization}
\label{sec:appendix_step_deviation}
To evaluate the accuracy of failure-step localization, we quantify the deviation between the true failure step and the step predicted by each method under the Qwen-7B and all-at-once settings on the Who\&When dataset. In Fig.~\ref{fig:step_difference}, the x-axis represents intervals of step prediction error, indicating how far the predicted failure step is from the actual one. The y-axis reports the number of failure cases that fall into each interval. For instance, in the Algorithm-Generated subset, \alg identifies the failure step within a 0-2 step deviation for 28 cases, whereas 3 cases fall within the 6-7 step deviation interval. The majority of failure cases in the Algorithm-Generated dataset lie within the 0-5 deviation range across all methods, reflecting relatively high accuracy in step prediction. The Hand-Crafted dataset is more difficult and contains many cases with deviations exceeding 30 steps. Even under this challenging setting, \alg produces the largest number of predictions within small deviation ranges (less than 5) and the fewest within larger deviation ranges (greater than 5), highlighting its stronger step-localization capability relative to the baseline approaches.

\section{Step-level accuracy under different tolerances}
\label{sec:appendix_step_tolerance}
In our default configuration, step-level accuracy measures the proportion of decisive error steps that are identified exactly. Following the evaluation protocol in~\citet{zhang2025agent}, we also examine a tolerance-based variant of this metric, in which predictions are counted as correct if the predicted step lies within a specified tolerance window around the true erroneous step. Fig.~\ref{fig:step_tolerance} presents the results on both the Algorithm-Generated and Hand-Crafted datasets using Qwen-7B under the all-at-once evaluation mode. Across all tolerance values, \alg consistently attains high step-level accuracy compared with baseline methods. On the Hand-Crafted dataset, \alg demonstrates a clear advantage, with accuracy increasing steadily as the tolerance window widens. On the Algorithm-Generated dataset, \alg remains highly competitive with AgenTracer and ECHO, confirming strong step-localization performance across different levels of difficulty and tolerance settings.

\section{Impact of trajectory length on localization accuracy}
\label{sec:appendix_token_length}

It is important to note that different failure cases may produce trajectories of varying lengths, measured by the total number of tokens. To analyze this effect, we group trajectories into four token-length categories. For instance, as shown in Fig.~\ref{fig:token_len}, in the Algorithm-Generated dataset, 31 failure cases have trajectories longer than 2000 tokens. In this Section, we investigate whether the length of a failure trajectory influences localization accuracy, focusing on the Qwen-7B model under the all-at-once mode. Table~\ref{dataset length level-agent} presents the agent-level accuracy of the various methods across the length categories, while Table~\ref{dataset length level-step} reports the corresponding step-level accuracy. In Table~\ref{dataset length level-step}, the AEGIS method is omitted because it does not provide step-level predictions.\\
As indicated in Table~\ref{dataset length level-agent} and Table~\ref{dataset length level-step}, neither the agent-level nor the step-level accuracy on the Algorithm-Generated dataset shows a clear monotonic relationship with token length. While all methods exhibit fluctuations across length ranges, \alg remains strong and comparatively stable, outperforming the baselines in most settings. On the Hand-Crafted dataset, precise step-level localization becomes particularly difficult for very long trajectories, where the true failure signal is diluted by lengthy interaction histories and substantial irrelevant context. Nevertheless, \alg still demonstrates notable robustness across diverse trajectory lengths, indicating that it is less affected by increasing sequence complexity and can produce reliable judgments even when failure logs differ substantially in size.

\section{Comparison between \alg and poisoning forensics approaches}
\label{sec:appendix_forensics}
Note that our work focuses on localizing system failures in multi-agent systems, where failures arise from incorrect reasoning, miscoordination, or erroneous intermediate decisions among agents. These failures are not the result of adversarial attacks, where maliciously injected or poisoned data intentionally manipulate model behavior. As a result, our method is fundamentally different from poisoning forensics approaches, which aim to trace back the specific components or data sources responsible for attack-induced misgeneration. In this paper, we further compare our \alg with state-of-the-art forensics methods, RAGOrigin~\citep{zhang2025taught} and RAGForensics~\citep{zhang2025traceback}. 
RAGOrigin uncovers harmful data by examining how each retrieved document influences the model’s produced answer. By selectively modifying or suppressing individual passages and observing the resulting changes in the LLM’s output, the method pinpoints the documents whose information most directly leads to an incorrect or malicious response.
RAGForensics carries out traceback by repeatedly retrieving a set of candidate documents and using an LLM classifier to determine whether each one contributes to the incorrect answer. Documents identified as poisoning sources are removed, and the procedure continues until only benign materials remain, thereby isolating the specific entries responsible for the erroneous response.
Both methods have demonstrated strong effectiveness in identifying poisoned data that lead to incorrect outputs in LLM-based systems, and we use their official implementations in our evaluation.

Table~\ref{results with RAG} presents a comparison between our \alg and these two approaches under both evaluation modes on the Algorithm-Generated dataset. The results show that poisoning forensics methods perform poorly in localizing system failures in multi-agent settings. This is because these approaches are specifically designed for poisoning forensics and are effective at capturing attack patterns, as adversarial perturbations are typically structured to influence model outputs and thus leave detectable traces, even when crafted to remain stealthy. In contrast, system failures in multi-agent systems stem from benign yet complex interactions, such as subtle reasoning errors or coordination breakdowns, which are difficult to isolate and do not exhibit the explicit signatures that poisoning forensics methods rely on.

\begin{table*}[t]
\centering
\small

\begin{tabular}{l lcccc}
\toprule
& & \multicolumn{2}{c}{All-at-once} & \multicolumn{2}{c}{Step-by-step} \\
\cmidrule(lr){3-4} \cmidrule(lr){5-6}
Model & Method & Agent-level & Step-level & Agent-level & Step-level \\
\midrule
\multirow{5}{*}{Qwen-7B}
& WhichAgent  & 45.83 & 4.17  & 33.33 & 4.17  \\
& AgenTracer  & 33.33 & 0.00  & 29.17 & 0.00  \\
& ECHO        & 8.33  & 8.33  & 4.17  & 4.17  \\
& AEGIS       & 54.17 & --    & 33.33 & --    \\
\rowcolor{greyL}
\cellcolor{white} & \alg & 62.50 & 16.67 & 50.00 & 37.50 \\
\midrule
\multirow{5}{*}{Llama-8B}
& WhichAgent  & 25.00 & 4.17  & 33.33 & 8.33  \\
& AgenTracer  & 16.67 & 4.17  & 41.67 & 0.00  \\
& ECHO        & 16.67 & 0.00 & 4.17  & 8.33  \\
& AEGIS       & 37.50 & --    & 45.83 & --    \\
\rowcolor{greyL}
\cellcolor{white} & \alg & 62.50 & 20.83 & 58.33 & 25.00 \\
\midrule
\multirow{5}{*}{Mistral-7B}
& WhichAgent  & 37.50 & 12.50 & 4.17  & 8.33  \\
& AgenTracer  & 16.67 & 0.00  & 4.17  & 0.00  \\
& ECHO        & 33.33 & 12.50 & 4.17  & 0.00  \\
& AEGIS       & 4.17  & --    & 12.50 & --    \\
\rowcolor{greyL}
\cellcolor{white} & \alg & 58.33 & 25.00 & 37.50 & 16.67 \\
\midrule
\multirow{5}{*}{GPT-4o}
& WhichAgent  & 58.33 & 8.33  & 25.00 & 8.33  \\
& AgenTracer  & 16.67 & 8.33  & 16.67 & 12.50 \\
& ECHO        & 16.67 & 0.00  & 4.17  & 4.17  \\
& AEGIS       & 29.17 & --    & 25.00 & --    \\
\rowcolor{greyL}
\cellcolor{white} & \alg & 54.17 & 16.67 & 50.00 & 25.00 \\
\bottomrule
\end{tabular}
\caption{Localization performance on the Who\&When \textit{Hand-Crafted} subset. Higher agent-level and step-level accuracy (\%) indicate better localization performance.}
\label{result_hand}
\end{table*}

\begin{table}[t]
\centering
\small
\addtolength{\tabcolsep}{-2.7pt}
\begin{tabular}{l lcccc}
\toprule

& &\multicolumn{2}{c}{All-at-once}  &\multicolumn{2}{c}{Step-by-step} \\
\cmidrule(lr){3-4} \cmidrule(lr){5-6}
Model & Method & Agent-level & Pair-level & Agent-level & Pair-level \\
\midrule
\multirow{4}{*}{Qwen-7B}
& WhichAgent   & 39.67 & 6.00 & 35.00 & 4.33 \\
& AgenTracer    & 46.17 & 7.33 & 42.67 & 6.67 \\
& AEGIS         & 39.83 & 6.67 & 41.00 & 6.17 \\
\rowcolor{greyL}
\cellcolor{white}& \alg    & 54.83 & 10.50 & 59.83 & 11.50 \\
\midrule
\multirow{4}{*}{Llama-8B}
& WhichAgent   & 43.00 & 4.33 & 41.50 & 5.33 \\
& AgenTracer    & 46.00 & 5.67 & 23.33 & 2.33 \\
& AEGIS         & 43.17 & 6.33 & 42.00 & 5.83 \\
\rowcolor{greyL}
\cellcolor{white}& \alg    & 52.83 & 7.17 & 55.67 &9.17 \\
\midrule
\multirow{4}{*}{Mistral-7B}
& WhichAgent  & 31.17 & 4.00 & 34.50 & 4.00 \\
& AgenTracer    & 43.17 & 3.67 & 26.33 & 3.67 \\
& AEGIS         & 34.50 & 4.33 & 33.50 & 3.17 \\
\rowcolor{greyL}
\cellcolor{white}& \alg    & 61.00 & 11.83 & 45.33 & 8.83 \\
\bottomrule
\end{tabular}
\caption{Localization performance of different methods on the Aegis-Bench dataset. Higher agent-level and pair-level accuracy (\%) indicate better localization performance.}
\label{aegis_result}
\end{table}

\begin{figure}[htbp]
    \centering
    \includegraphics[width=1\textwidth]{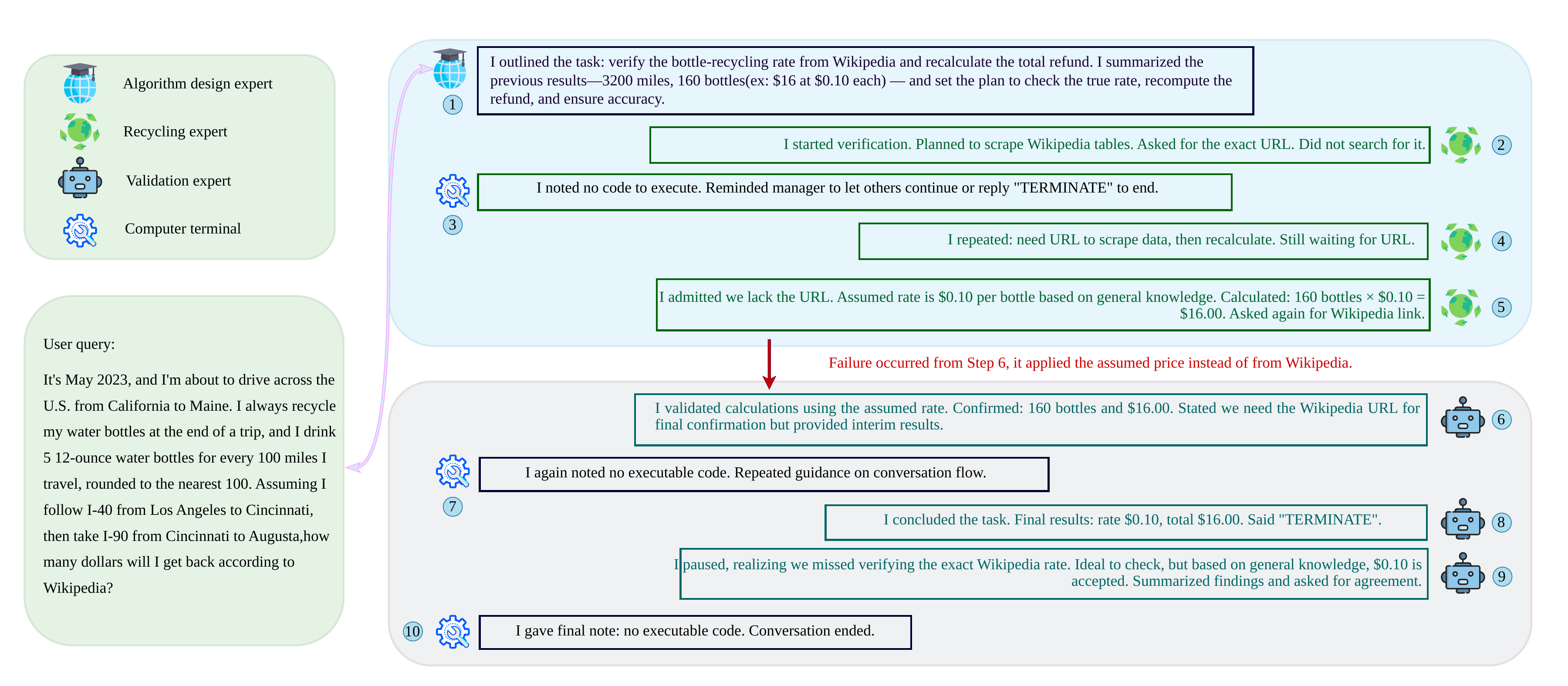}
     \caption{Illustrative example of failure propagation in a multi-agent system.}
    \label{fig:Failure propagation}
    \vspace{-.10in}
\end{figure} 

\begin{figure}[t]
    \centering
    \includegraphics[width=0.9\textwidth]{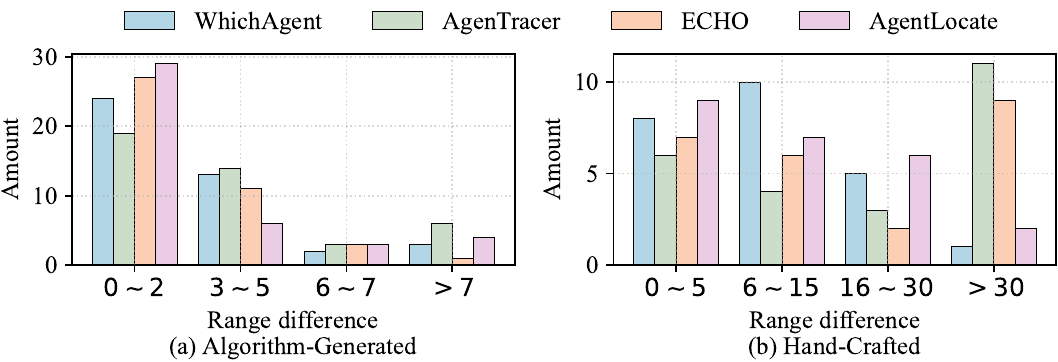}
    \caption{Distribution of step localization error.}
    \label{fig:step_difference}
    \vspace{-.10in}
\end{figure}

\begin{figure}[t]
    \centering
    \includegraphics[width=0.9\textwidth]{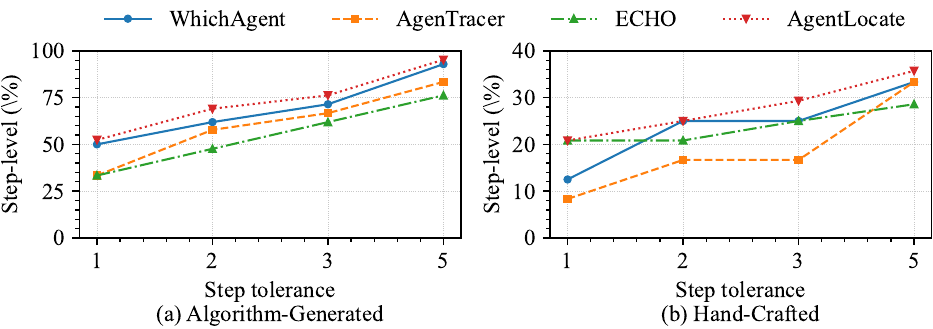}
     \caption{Localization performance of different methods with different step tolerance parameters.}
    \label{fig:step_tolerance}
    \vspace{-.10in}
\end{figure}

\begin{figure}[t]
    \centering
    \includegraphics[width=0.9\textwidth]{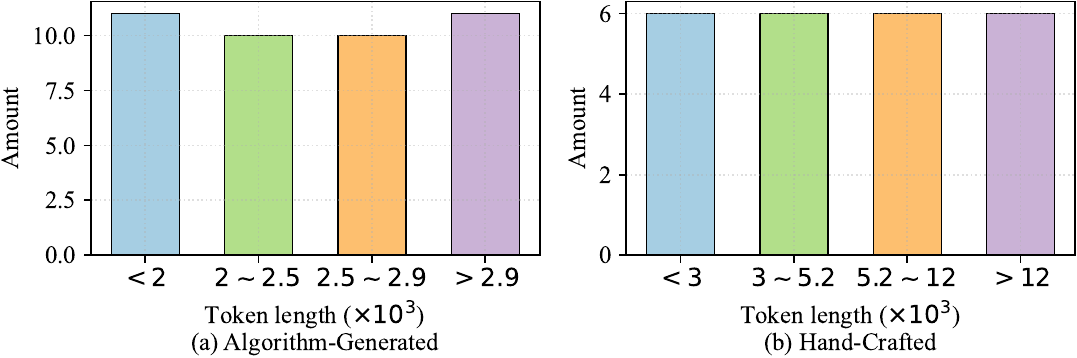}
    \caption{Distribution of failure trajectories by token length.}
    \label{fig:token_len}
    \vspace{-.10in}
\end{figure}

\begin{table}[t]
\centering
\small
\addtolength{\tabcolsep}{-2.7pt}
\renewcommand{\arraystretch}{1.2}
\begin{tabular}{lcccccccc}
\toprule
 & \multicolumn{4}{c}{Algorithm-Generated} & \multicolumn{4}{c}{Hand-Crafted} \\
\cmidrule(lr){2-5} \cmidrule(lr){6-9}
Length ($\times10^3$) & $<2$ & $2\!\sim\!2.5$ & $2.5\!\sim\!2.9$ & $>2.9$
                 & $<3$ & $3\!\sim\!5.2$ & $5.2\!\sim\!12$ & $>12$ \\
\midrule
WhichAgent   & 45.45 & 40.00 & 50.00 & 18.18 & 50.00 & 50.00 & 33.33 & 50.00 \\
AgenTracer   & 36.36 & 40.00 & 50.00 & 54.55 & 33.33 & 33.33 & 16.67 & 50.00 \\
ECHO         & 45.45 & 50.00 & 30.00 & 0.00 & 16.67 & 0.00  & 16.67 & 0.00  \\
AEGIS        & 36.36 & 40.00 & 30.00 & 27.27 & 50.00 & 66.67 & 66.67 & 50.00 \\
\rowcolor{greyL}
\alg         & 54.55 & 70.00 & 80.00 & 72.73 & 66.67 & 83.33 & 66.67 & 33.33 \\
\bottomrule
\end{tabular}
\caption{Agent-level accuracy (\%) across different trajectory token-length ranges.}
\label{dataset length level-agent}
\end{table}

\begin{table}[t]
\centering
\small
\addtolength{\tabcolsep}{-2.7pt}
\renewcommand{\arraystretch}{1.2}
\begin{tabular}{lcccccccc}
\toprule
 & \multicolumn{4}{c}{Algorithm-Generated} & \multicolumn{4}{c}{Hand-Crafted} \\
\cmidrule(lr){2-5} \cmidrule(lr){6-9}
Length ($\times10^3$) & $<2$ & $2\!\sim\!2.5$ & $2.5\!\sim\!2.9$ & $>2.9$
                 & $<3$ & $3\!\sim\!5.2$ & $5.2\!\sim\!12$ & $>12$ \\
\midrule
WhichAgent   & 18.18 & 20.00 & 10.00 & 27.27 & 16.67 & 0.00  & 0.00  & 0.00  \\
AgenTracer   & 18.18 & 10.00 & 10.00 & 9.09  & 0.00  & 0.00  & 0.00  & 0.00  \\
ECHO         & 9.09  & 20.00 & 30.00 & 0.00  & 16.67 & 0.00  & 16.67 & 0.00  \\
\rowcolor{greyL}
\alg         & 36.36 & 30.00 & 40.00 & 45.45 & 50.00 & 16.67 & 0.00  & 0.00  \\
\bottomrule
\end{tabular}
\caption{Step-level accuracy (\%) across different trajectory token-length ranges.}
\label{dataset length level-step}
\end{table}

\begin{table}[t]
\centering
\small
\addtolength{\tabcolsep}{-0.7pt}
\begin{tabular}{lcccc}
\toprule
 & \multicolumn{2}{c}{Algorithm-Generated} & \multicolumn{2}{c}{Hand-Crafted} \\
\cmidrule(lr){2-3}\cmidrule(lr){4-5}
 Method & All-at-once & Step-by-step & All-at-once & Step-by-step \\
\midrule
WhichAgent & 248,662 & 589,638 & 494,274 & 1,017,966 \\
AgenTracer & 16,419,207 & 16,451,873 & 23,072,290 & 23,161,747 \\
ECHO & 521,642 & 591,916 & 608,504 & 1,789,149 \\
\rowcolor{greyL}
\alg & 451,776 & 574,817 & 1,071,606 & 1,294,065 \\
\bottomrule
\end{tabular}
\caption{Number of tokens used by different methods (Qwen-7B).}
\label{tab:token_cost_tokens}
\end{table}

\begin{table}[t]
\centering
\small
\addtolength{\tabcolsep}{-0.7pt}
\begin{tabular}{lcccc}
\toprule
 & \multicolumn{2}{c}{Algorithm-Generated} & \multicolumn{2}{c}{Hand-Crafted} \\
\cmidrule(lr){2-3}\cmidrule(lr){4-5}
 Method & All-at-once & Step-by-step & All-at-once & Step-by-step \\
\midrule
WhichAgent & 0.034 & 0.082 & 0.068 & 0.141 \\
AgenTracer & 2.275 & 2.278 & 3.204 & 3.210 \\
ECHO       & 0.072 & 0.082 & 0.084 & 0.248 \\
\rowcolor{greyL}
\alg       & 0.063 & 0.080 & 0.148 & 0.180 \\
\bottomrule
\end{tabular}
\caption{The monetary cost (in USD) of different methods (Qwen-7B).}
\label{tab:token_cost_money}
\end{table}

\begin{table}[t]
\centering
\small
\addtolength{\tabcolsep}{-0.7pt}

\begin{tabular}{lcccc}
\toprule
 & \multicolumn{2}{c}{Algorithm-Generated} & \multicolumn{2}{c}{Hand-Crafted} \\
\cmidrule(lr){2-3}\cmidrule(lr){4-5}
Method & All-at-once & Step-by-step & All-at-once & Step-by-step \\
\midrule
WhichAgent   & 260 & 487 & 556 & 1,835 \\
AgenTracer    & 6,809 & 6,822  & 6,752  & 6,776 \\
ECHO          & 212 & 449 & 349  & 550 \\
\rowcolor{greyL}
\alg   & 317  & 355  & 315   & 338 \\
\bottomrule
\end{tabular}
\caption{Running time (seconds) of different methods (Qwen-7B).}
\label{tab:token_cost_time}
\end{table}

\begin{table*}[t]
\centering
\small
\subfloat[Algorithm-Generated]{
\begin{tabular}{lcccc}
\toprule
& \multicolumn{2}{c}{All-at-once} & \multicolumn{2}{c}{Step-by-step} \\
\cmidrule(lr){2-3} \cmidrule(lr){4-5}
Model & Agent-level & Step-level & Agent-level & Step-level \\
\midrule
Qwen-7B    & 69.05 & 38.10 & 69.05 & 33.33 \\
Llama-8B   & 57.14 & 28.57 & 57.14 & 23.81 \\
Mistral-7B & 52.38 & 23.81 & 50.00 & 14.29 \\
GPT-4o     & 61.90 & 30.95 & 57.14 & 26.19 \\
\bottomrule
\end{tabular}
}
\hfill
\subfloat[Hand-Crafted]{
\begin{tabular}{lcccc}
\toprule
& \multicolumn{2}{c}{All-at-once} & \multicolumn{2}{c}{Step-by-step} \\
\cmidrule(lr){2-3} \cmidrule(lr){4-5}
Model & Agent-level & Step-level & Agent-level & Step-level \\
\midrule
Qwen-7B    & 62.50 & 16.67 & 50.00 & 37.50 \\
Llama-8B   & 62.50 & 20.83 & 50.00 & 25.00 \\
Mistral-7B & 54.17 & 12.50 & 54.17 & 16.67 \\
GPT-4o     & 58.33 & 16.67 & 50.00 & 29.17 \\
\bottomrule
\end{tabular}
}
\caption{Localization accuracy (\%) of \alg under different evaluator models, with the Judge fixed to Qwen-7B: (a) Algorithm-Generated subset. (b) Hand-Crafted subset.}
\label{results with different evaluator model}
\end{table*}

\begin{table*}[t]
\centering
\small
\subfloat[All-at-once]{
\begin{tabular}{lcccccccc}
\toprule
& \multicolumn{4}{c}{Agent-level} & \multicolumn{4}{c}{Step-level} \\
\cmidrule(lr){2-5} \cmidrule(lr){6-9}
Model & 1 & 3 & 5 & 7 & 1 & 3 & 5 & 7 \\
\midrule
Qwen-7B    & 35.71 & 69.05 & 52.38 & 66.67 & 16.67 & 38.10 & 21.43 & 35.71 \\
Llama-8B   & 47.62 & 54.76 & 52.38 & 50.00 & 19.05 & 23.81 & 30.95 & 16.67 \\
Mistral-7B & 47.62 & 50.00 & 54.76 & 57.14 & 23.81 & 26.19 & 26.19 & 30.95 \\
\bottomrule
\end{tabular}
}
\hfill
\subfloat[Step-by-step]{
\begin{tabular}{lcccccccc}
\toprule
& \multicolumn{4}{c}{Agent-level} & \multicolumn{4}{c}{Step-level} \\
\cmidrule(lr){2-5} \cmidrule(lr){6-9}
Model & 1 & 3 & 5 & 7 & 1 & 3 & 5 & 7 \\
\midrule
Qwen-7B    & 35.71 & 69.05 & 64.29 & 69.05 & 16.67 & 33.33 & 21.43 & 33.33 \\
Llama-8B   & 47.62 & 54.76 & 52.38 & 50.00 & 19.05 & 16.67 & 21.43 & 19.05 \\
Mistral-7B & 35.71 & 38.10 & 47.62 & 54.76 & 16.67 & 16.67 & 23.81 & 33.33 \\
\bottomrule
\end{tabular}
}
\caption{Localization accuracy (\%) of \alg with different numbers of Evaluators: (a) All-at-once mode. (b) Step-by-step mode.}
\label{results with different evaluator number}
\end{table*}

\begin{table*}[htbp]
\centering
\addtolength{\tabcolsep}{-1.55pt}
\subfloat[All-at-once]{
\begin{tabular}{lcccccc}
\toprule
& \multicolumn{3}{c}{Agent-level} & \multicolumn{3}{c}{Step-level} \\
\cmidrule(lr){2-4} \cmidrule(lr){5-7}
Model & 1 round & 2 rounds & 3 rounds & 1 round & 2 rounds & 3 rounds \\
\midrule
Qwen-7B    & 69.05 & 71.43 & 69.05 & 38.10 & 38.10 & 38.10 \\
Llama-8B   & 54.76 & 57.14 & 54.76 & 23.81 & 23.81 & 26.19 \\
Mistral-7B & 50.00 & 52.38 & 52.38 & 26.19 & 23.81 & 23.81 \\
\bottomrule
\end{tabular}
}
\hfill
\subfloat[Step-by-step]{
\begin{tabular}{lcccccc}
\toprule
& \multicolumn{3}{c}{Agent-level} & \multicolumn{3}{c}{Step-level} \\
\cmidrule(lr){2-4} \cmidrule(lr){5-7}
Model & 1 round & 2 rounds & 3 rounds & 1 round & 2 rounds & 3 rounds \\
\midrule
Qwen-7B    & 69.05 & 69.05 & 66.67 & 33.33 & 28.57 & 33.33 \\
Llama-8B   & 54.76 & 57.14 & 57.14 & 16.67 & 19.05 & 21.43 \\
Mistral-7B & 38.10 & 42.86 & 47.62 & 16.67 & 21.43 & 21.43 \\
\bottomrule
\end{tabular}
}
\caption{Impact of different numbers of Judge-Evaluator refinement rounds on the localization accuracy (\%) of \alg on the Algorithm-Generated subset: (a) All-at-once mode. (b) Step-by-step mode.}
\label{results_loop_algorithm}
\end{table*}

\begin{table*}[htbp]
\centering
\addtolength{\tabcolsep}{-1.55pt}
\subfloat[All-at-once]{
\begin{tabular}{lcccccc}
\toprule
& \multicolumn{3}{c}{Agent-level} & \multicolumn{3}{c}{Step-level} \\
\cmidrule(lr){2-4} \cmidrule(lr){5-7}
Model & 1 round & 2 rounds & 3 rounds & 1 round & 2 rounds & 3 rounds \\
\midrule
Qwen-7B    & 62.50 & 62.50 & 62.50 & 16.67 & 25.00 & 29.17 \\
Llama-8B   & 62.50 & 58.33 & 62.50 & 20.83 & 25.00 & 25.00 \\
Mistral-7B & 58.33 & 58.33 & 58.33 & 25.00 & 25.00 & 25.00 \\
\bottomrule
\end{tabular}
}
\hfill
\subfloat[Step-by-step]{
\begin{tabular}{lcccccc}
\toprule
& \multicolumn{3}{c}{Agent-level} & \multicolumn{3}{c}{Step-level} \\
\cmidrule(lr){2-4} \cmidrule(lr){5-7}
Model & 1 round & 2 rounds & 3 rounds & 1 round & 2 rounds & 3 rounds \\
\midrule
Qwen-7B    & 50.00 & 50.00 & 54.17 & 37.50 & 33.33 & 33.33 \\
Llama-8B   & 58.33 & 58.33 & 58.33 & 25.00 & 25.00 & 29.17 \\
Mistral-7B & 37.50 & 41.67 & 41.67 & 16.67 & 25.00 & 20.83 \\
\bottomrule
\end{tabular}
}
\caption{Impact of different numbers of Judge-Evaluator refinement rounds on the localization accuracy (\%) of \alg on the Hand-Crafted subset: (a) All-at-once mode. (b) Step-by-step mode.}
\label{results_loop_handcrafted}
\end{table*}

\begin{table*}[t]
\centering
\subfloat[Algorithm-Generated]{
\begin{tabular}{lcccc}
\toprule
& \multicolumn{2}{c}{All-at-once} & \multicolumn{2}{c}{Step-by-step} \\
\cmidrule(lr){2-3} \cmidrule(lr){4-5}
Variant & Agent-level & Step-level & Agent-level & Step-level \\
\midrule
Variant I   & 38.10 & 19.05 & 38.10 & 16.67 \\
Variant II  & 52.38 & 28.57 & 47.62 & 23.81 \\
Variant III & 54.76 & 23.81 & 57.14 & 26.19 \\
\rowcolor{greyL}
\alg        & 69.05 & 38.10 & 69.05 & 33.33 \\
\bottomrule
\end{tabular}
}
\hfill
\subfloat[Hand-Crafted]{
\begin{tabular}{lcccc}
\toprule
& \multicolumn{2}{c}{All-at-once} & \multicolumn{2}{c}{Step-by-step} \\
\cmidrule(lr){2-3} \cmidrule(lr){4-5}
Variant & Agent-level & Step-level & Agent-level & Step-level \\
\midrule
Variant I   & 29.17 & 16.67 & 37.50 & 12.50 \\
Variant II  & 45.83 & 12.50 & 41.67 & 20.83 \\
Variant III & 50.00 & 16.67 & 41.67 & 25.00 \\
\rowcolor{greyL}
\alg        & 62.50 & 16.67 & 50.00 & 37.50 \\
\bottomrule
\end{tabular}
}
\caption{Localization accuracy (\%) of different variants of \alg: (a) Algorithm-Generated subset. (b) Hand-Crafted subset.}
\label{results with variants}
\end{table*}

\begin{table}[htbp]
\centering
\addtolength{\tabcolsep}{-1.7pt}
\begin{tabular}{lcccc}
\toprule
\multirow{2}{*}{{}} 
& \multicolumn{2}{c}{{All-at-once}} 
& \multicolumn{2}{c}{{Step-by-step}} \\
\cmidrule(lr){2-3} \cmidrule(lr){4-5}
Model & Agent-level & Step-level & Agent-level & Step-level \\
\midrule
RAGOrigin    & 28.57 & 9.52 & 28.57 & 16.67 \\
RAGForensics & 30.95 & 9.52  & 33.33 & 19.05 \\
\rowcolor{greyL}
\alg  & 69.05 & 38.10 & 69.05 &33.33 \\
\bottomrule
\end{tabular}
\caption{Comparison of \alg with poisoning forensics approaches.}
\label{results with RAG}
\end{table}

\begin{table}[t]
\centering
\small
\begin{tabular}{lcc}
\toprule
{Method} & Algorithm-Generated & Hand-Crafted\\
\midrule
WhichAgent   & Verification\_Expert   & WebSurfer  \\
AgenTracer   & DataAnalysis\_Expert & WebSurfer  \\
ECHO         & Computer\_Terminal   & WebSurfer  \\
AEGIS        & Verification\_Expert   & WebSurfer \\
\rowcolor{greyL}
\alg   & Verification\_Expert & Assistant \\
\bottomrule
\end{tabular}
\caption{The most frequently mislocalized agents for different methods.}
\label{Top 3 error agent}
\end{table}

\begin{figure}[t]
    \centering
    \includegraphics[width=0.9\textwidth]{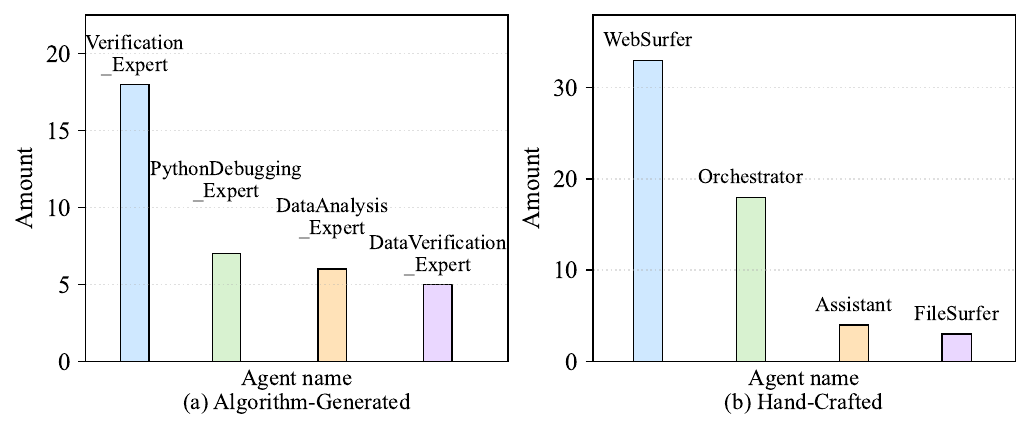}
  \caption{Top-4 most frequent failure-causing agents in the Who\&When dataset.}
    \label{fig:error agent freq}
    \vspace{-.10in}
\end{figure}

\end{document}